\documentclass[onecolumn,pra]{revtex4}
\usepackage{amsmath}
\usepackage{graphicx}
\usepackage{amssymb}

\begin{document}

\title{Imaging mesoscopic nuclear spin noise with a diamond magnetometer }

\author{Carlos A. Meriles$^*$, Liang Jiang$^{2,}$\footnote{\ Equally contributing authors.}\footnote{ Present 
Address: Institute for Quantum Information, Caltech, Pasadena, CA 91125, USA.},
Garry Goldstein$^{2}$,  Jonathan~S.~Hodges$^{2,3}$\footnote{ Present Address: Department of Electrical 
Engineering, Columbia University, New York, NY 10023, USA.}, Jeronimo R. Maze$^{2}$,
 Mikhail D. Lukin$^{2}$ and Paola Cappellaro$^{3}$,}
\affiliation{$^{1}$Department of Physics, The City College of New York, CUNY New York, NY 10031, USA\\  $^{2}$Department of Physics, Harvard University, Cambridge, MA 02138, USA \\ $^3$ Department of Nuclear Science and Engineering, Massachusetts Institute of Technology, Cambridge, MA 02139, USA}

\begin{abstract}
Magnetic Resonance Imaging (MRI) can characterize and discriminate among tissues 
using their diverse physical and biochemical properties. Unfortunately, submicrometer 
screening of biological specimens is presently not possible, mainly due to lack of 
detection sensitivity. Here we analyze the use of a nitrogen-vacancy center in diamond 
as a magnetic sensor for nanoscale nuclear spin imaging and spectroscopy. We examine 
the ability of such a sensor to probe the fluctuations of the `classical' dipolar 
field due to a large number of neighboring nuclear spins in a densely protonated 
sample. We identify detection protocols that appropriately take into account the 
quantum character of the sensor and find a signal-to-noise ratio compatible with 
realistic experimental parameters. Through various example calculations we illustrate 
different kinds of image contrast. In particular, we show how to exploit the comparatively 
long nuclear spin correlation times to reconstruct a local, high-resolution sample 
spectrum. 
\end{abstract}
\maketitle

 \section{ Introduction}

Physical tools have historically facilitated advances in biology; notable examples 
are X-rays crystallography, DNA sequencing, microarrays techniques, and, above all, 
microscopy in its various forms. Extending Nuclear Magnetic Resonance (NMR) to the 
micro- and nano-scale promises to become another leading resource in the microscopist's 
toolbox: Unlike any other technique, NMR is unique in allowing the generation of 
images with different information content. Multidimensional high-resolution spectroscopy 
is today routinely used in the liquid and solid state to unveil complex molecular 
structures, and this capability could prove groundbreaking if samples having sub-microscopic 
dimensions could be efficiently probed. Unfortunately, these features cannot be fully 
exploited at present because NMR lacks the sensitivity essential to high-resolution 
screening. The origin of this limitation is twofold: First, in ``conventional'' NMR 
the signal to noise ratio ($SNR$) is proportional to the nuclear magnetic 
polarization of the sample, which represents only a small fraction of the attainable 
maximum ($\sim10^{-4}$ for protons in a 14 T magnet at 300 K). Second, Faraday induction 
is a poor detection method since, even with maximum polarization, the minimum number 
of spins needed to induce a measurable signal is comparatively large.

Although experiments performed at lower temperatures and/or higher fields can partly 
mitigate these problems, other more efficient detection techniques have recently 
been proposed. One strategy is to use the spin associated to a single nitrogen-vacancy 
(NV) center in diamond as a local magnetic field probe.\cite{Taylor08,Degen08}. The operating principles 
of this approach closely mimic those of an atomic vapor magnetometer~\cite{Budker07}, where the applied 
magnetic field is inferred from the shift in the Larmor precession frequency. Owing 
to the exceptionally long coherence times of NV centers---exceeding 1 ms at room 
temperature in ultra-pure bulk samples~\cite{Balasubramanian09}---detection of 
3 nT over a measurement time of only 100 s has been experimentally demonstrated.\cite{Maze08} Further, a NV center within a diamond nanocrystal attached to an AFM tip 
was recently used to image a magnetic nanostructure with 20 nm resolution.\cite{Balasubramanian08} 

Here we focus on applications of a NV center mounted on a scanning probe for monitoring 
adjacent nuclear spins in an external, infinitely-extended organic sample. Rather 
than detecting single nuclear spins---an extremely challenging goal---we focus on 
the case where the NV center interacts with large ensembles of nuclear spins localized 
over effective volumes of $\sim$(10-50nm)$^3$. This regime lends itself to a simplified 
description that simultaneously takes into consideration the quantum nature of the 
sensor---the NV center---while relying on a classical description of the long-range 
dipolar fields induced by the nuclear spin ensemble. Similar to prior magnetic resonance 
force microscopy experiments\cite{Degen07}, our strategy exploits the small dimensions of the 
effective sample to probe the `nuclear spin noise'---i.e., the statistical fluctuations 
of the nuclear magnetization---rather than the magnetization itself. An important 
consequence is that, unlike traditional MRI, spatial resolution is not due to strong 
magnetic field gradients but is rather determined by the distance between the NV 
center and the sample. Assuming a very small external magnetic field we determine 
the conditions required for 2D nuclear spin imaging at (or near) room temperature, 
and show them to be compatible with realistic experimental parameters. Further, we 
show that, in addition to determining the local nuclear spin density, this strategy 
allows one to explore different kinds of contrast mechanisms (nearly a requisite 
when imaging, for example, densely protonated organic/biological systems). In particular, 
we show how to reconstruct the local nuclear spin correlation function and, from 
it, a spatially-resolved nuclear spin spectrum. 

 The paper is organized as follows. First, we briefly review the operating principles 
of NV-center-based magnetometry, more explicitly identify the effective size of the 
sample being probed, and lay out our detection protocol. Subsequently, we describe 
different modalities of `nuclear spin noise' detection and determine in each case 
the limit signal-to-noise ratio. Finally, we discuss image contrast and localized 
nuclear spin spectroscopy and conclude with some model calculations.

  \section{Spin-noise magnetometry with a single NV center}

The negatively charged nitrogen-vacancy center in diamond is an impurity comprising 
a total of six electrons, two of which are unpaired and form a triplet ground state 
with a zero-field splitting $D_{gs}$=2.87 GHz. In our calculations we assume 
the presence of a small magnetic field $B_{A} \hat{z}$  ($\sim10$ mT) collinear 
with the crystal field (which, in turn, is oriented either along the [111] axis or 
its crystallographic equivalents). Though non-mandatory, the auxiliary field lifts 
the degeneracy between the ${\left| m_{s} =\pm 1 \right\rangle} $ states, thus allowing 
one to selectively address only one of the two possible transitions, e.g., between ${
\left| m_{s} =0 \right\rangle} $ and ${\left| m_{s} =1 \right\rangle} $. 

When a green laser (532 nm) illuminates the NV center, the system is excited into 
an optically active triplet state; subsequent intersystem crossing produces a dark, 
singlet state that preferentially relaxes into ${\left| m_{s} =0 \right\rangle} $. 
Almost complete optical pumping of the ground state takes place after a $\sim1\mu$s 
illumination, thus allowing us to model the initial density matrix of the NV center --for 
practical purposes, a two-level system-- as 

\begin{equation} \label{Eq:1} 
\rho \left(0\right)={\left| 0 \right\rangle} {\left\langle 0 \right|} =\frac{1}{2} 
\left(I+\sigma _{z} \right),      
\end{equation} 
where $I$ denotes the identity operator and $\sigma _{z} $ is the Pauli matrix. 
Because intersystem crossing is allowed only if excitation takes place from ${\left| 
m_{s} =1 \right\rangle} $, the fluorescence intensity correlates with the population 
of the spin state.  We model the `measurement' operator as

\begin{equation} \label{Eq:2} 
M=a{\left| 0 \right\rangle} {\left\langle 0 \right|} +b{\left| 1 \right\rangle} {
\left\langle 1 \right|} =\frac{1}{2} \left(a+b\right)I+\frac{1}{2} \left(a-b\right)
\sigma _{z} .   
\end{equation} 
In Eq. (\ref{Eq:2}), $a$ and $b$ are two independent, stochastic 
variables associated with the  total number of photons collected during the measurement 
interval ($\sim300$ ns) and characterized by Poisson distributions 
$q_{a} (k)=\alpha^{k} e^{-\alpha }/k!$ and $q_{b} (k)=\beta ^{k} e^{-\beta } /k!$, with $k$ integer. Due to the branching ratio into the dark singlet level, 
the averages over several measurements $\alpha \equiv \left\langle a\right\rangle $ and $\beta 
\equiv \left\langle b\right\rangle $ are substantially different ($\alpha \cong 1.5
\beta $) and thus provide the contrast necessary to discriminate the sensor spin 
state. 

Fig. 1a schematically shows the basics of our detection protocol: spin initialization 
and a selective $\pi/2$ microwave pulse are followed by a period $\Delta t$ of free 
evolution in the presence of an unknown, nuclear-spin-induced magnetic field $B_{N} 
\hat{z}$. Preceding optical readout, a second $\pi/2$ pulse, shifted by a phase $\theta$ relative 
to the first pulse, partially converts spin coherence into population differences. 
In the rotating frame resonant with the chosen transition, the density matrix describing 
the NV center is given by 

\begin{equation} \label{Eq:3} 
\rho \left(\Delta t\right)=\frac{1}{2} \left(I-\sigma _{x} \sin \left(\phi +\theta 
\right)+\sigma _{z} \cos \left(\phi +\theta \right)\right),    
\end{equation} 
where $\phi =\int _{0}^{\Delta t}\gamma _{e} B_{N} \left(t\right)\, dt $ denotes 
the total accumulated phase due to the nuclear field and $\gamma _{e} $ is the electronic 
gyromagnetic ratio. As in any other magnetometer-based strategy, the goal of a measurement 
is to extract the value of $\phi $ and, from it, valuable information on the magnetic 
field. 

\begin{figure}[htb]
	\centering
		\includegraphics[scale=0.25]{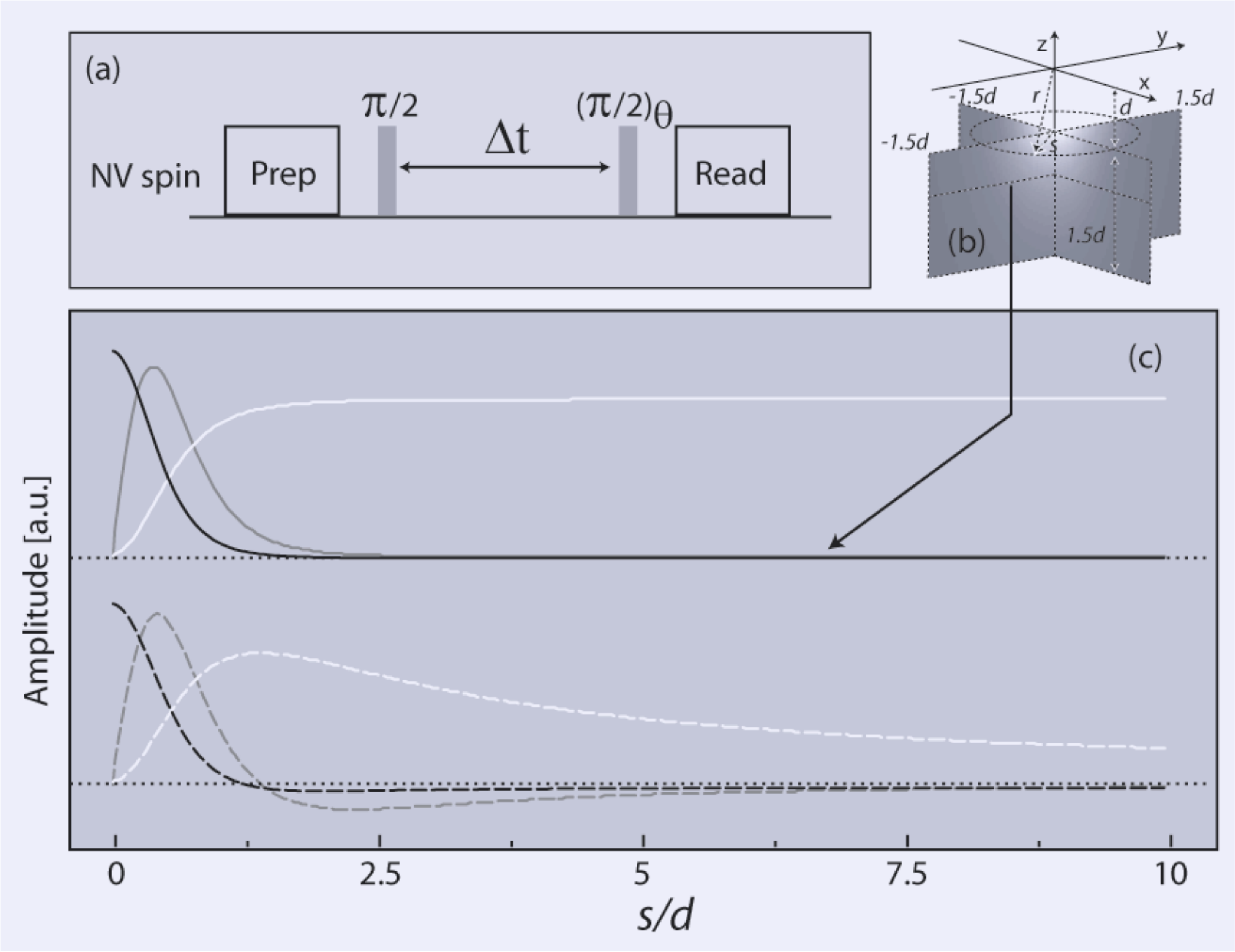}
	\caption{(a) Basic diamond-based magnetometry pulse sequence. (b) 
With the NV center at the reference frame origin, the grayscale indicates the relative 
contribution to field fluctuations from spins in a uniformly dense film. (c) In units 
of the relative radial coordinate $s/d$, the upper set of curves shows a cross 
section of the graph in (b) (black curve) and the corresponding integral (white curve). 
The grey curve shows the effective spin noise `density' $b_{N} \left(s,z\right)$ (see 
text). For comparison, the lower set shows the same curves but for the \textit{average} field 
at the NV center. Note that the integral (dashed white curve) decays slowly to zero 
as a result of negative contributions from spins far from the center. }
	\label{fig:Fig1}
\end{figure}

Before considering the constraints deriving from the quantum character of the sensor, 
we describe the magnetic field generated by the nuclear spin ensemble. In an experimental 
setup where the NV center scans an infinitely-extended sample film, the electronic 
sensor spin and the nuclear spins are coupled via long-range dipolar interactions. 
Given that in the rotating frame resonant with the sensor spin only components of 
the nuclear field parallel to the $z-$axis need be taken into consideration, 
we find 
\begin{equation} \label{Eq:4} 
\vec{{\bf B}}_{N} =B_{N} \hat{z}=\sum _{i}\left\{f\left(\vec{r}_{i} \right)\, \vec{m}_{z}^{
\left(i\right)} +g\left(\vec{r}_{i} \right)\, \left(\vec{m}_{\bot }^{\left(i\right)} \cdot 
\hat{r}_{i} \right)\hat{z}\right\} .   
\end{equation} 
Here $f\left(\vec{r}\right)=\left(\mu _{0}/4\pi r^{3}  \right)\left(3\cos 
^{2} \vartheta -1\right)$ and $g\left(\vec{r}\right)=\left(3\mu _{0}/4\pi r^{3}\right)\cos\vartheta $ are functions of the distance $r_{i} =\left|\vec{r}_{i} \right|$ of 
the $i$-th nuclear spin to the NV center and $\vartheta _{i} $ is the angle between 
the position vector and the $z-$axis; $\mu _{0} $ is the magnetic permeability 
of vacuum, and $\vec{m}_{z}^{\left(i\right)} $ ($\vec{m}_{\bot }^{\left(i\right)} $) 
denotes the component of the corresponding nuclear magneton $\vec{m}^{\left(i\right)} $ parallel 
(perpendicular) to the $z$-axis. We will consider the situation where the 
distance $d$ between the sensor and the surface is of order $\sim10$ nm or greater. 
We also assume that nuclear spins are dense (i.e., no nuclear spin can be singled 
out). In this regime, the NV center interacts with a large number of protons --exceeding 
$10^5$ in most organic samples-- and thus exerts a negligible back-action on the sample 
system. Each nuclear spin can be described classically via a stochastic, ergodic 
variable featuring first and second moments $\left\langle \vec{m}\right\rangle $ and $\left
\langle \vec{m}^{2} \right\rangle $, respectively.  

To see that detection of the \textit{time-dependent fluctuations }of the nuclear 
field---rather than the field itself---better suits our purpose, let us consider 
the case of a \textit{uniformly magnetized film} and assume, for simplicity, that 
the normal to the sample surface coincides with the $z$-axis. Using (\ref{Eq:4}) 
we write the time-averaged field acting on the sensor as 

\begin{equation} \label{Eq:5} 
\left\langle B_{N} \right\rangle =\frac{1}{V_{p} } \int _{Film}\left\{f\left(\vec{r}
\right)\, \left\langle m_{z} \right\rangle +g\left(\vec{r}\right)\, \left(\left\langle 
\vec{m}_{\bot } \right\rangle \cdot \hat{r}\right)\right\}\, dV ,    
\end{equation} 
where we have transformed the sums into volume integrals via the correspondence $\sum 
_{i}\to \;  \int {\tfrac{dV}{V_{p} }}  $ with $V_p$ representing the volume 
of the `primitive cell' associated to a single nuclear spin. From symmetry considerations, 
we observe that the second term in (\ref{Eq:5}) cancels out. This is also 
the case for the first term --in agreement with the classical magnetostatics result 
outside a thin, infinitely-extended, uniformly-polarized film-- but here a more subtle 
balance between contributions from spins close and far away from the sensor is responsible~\cite{Meriles05}. 
The latter is shown in Fig.~1c where we plot $f\left(\vec{r}\right)$ (and its integral) as a function of the (normalized) radial 
coordinate $s$ on the sample plane; within each thin slice of thickness $dz$, 
long-range, weaker contributions from more numerous spins far from the sensor exactly 
cancel the field created by spins contained within a central disk (of diameter comparable 
to the sensor-slice distance). 

 The concept of spin noise detection capitalizes on the spontaneous fluctuations 
of the nuclear spin magnetization in a small volume. To more quantitatively identify 
the sample volume within the film, consider the special case of a uniformly distributed, 
infinitely-extended sample and calculate the nuclear field variance $\Delta B_{N} 
^{2} $. Starting from Eq. (\ref{Eq:4}) and in the limit of (\ref{Eq:5}) 
we find 
\begin{equation} \label{Eq:6} 
\Delta B_{N} ^{2} =\left\langle B_{N} ^{2} \right\rangle \cong \frac{1}{V_{p} } \int_{Film}\left\{\left(f\left(\vec{r}\right)\right)^{2} \left\langle m_{z}^{2} \right
\rangle +\left(g\left(\vec{r}\right)\sin \vartheta \right)^{2} \left\langle m_{\bot 
}^{2} \right\rangle \right\}\, dV ,  
\end{equation} 
where we assumed $\left\langle m_{x}^{2} \right\rangle =\left\langle m_{y}^{2} \right
\rangle =\left\langle m_{\bot }^{2} \right\rangle $. 
Using cylindrical coordinates for convenience, we plot in Fig.~1c the `spin noise 
density',  
$b_{N}(s,z)\, ds\, dz=\frac{1}{V_{p} } \int _{0}^{2\pi }
\left\{\left[\left(f(s,z)\right)^{2} \langle m_{z}^2 \rangle +
\left(g(s,z)s/\sqrt{z^{2} +s^{2}}\right)^2 \langle m_{\bot }^{2} 
\rangle \right]\, s\, d\varphi \right\}\,  \, ds\, dz$. While spins far from the 
sensor have a non-negligible contribution, fluctuations of the nuclear field at the 
NV center are dominated by spins approximately contained within half a sphere of 
radius comparable to the sensor-surface distance $d$. Comparing with the prior 
results, we conclude that fluctuations selectively highlight spins close to the sensor --as 
opposed to `distant' spins-- not because the resulting \textit{average field }is 
stronger but because, being less numerous, the \textit{relative field variance} is 
larger. 

 A practical upper limit for the NV center-sample distance $d$ stems from 
the fact that the amplitude of the field fluctuations decreases sharply with the 
sensor-sample distance: Assuming a sample with spin density $\rho_N\sim 1/V_p$, 
we find 
\begin{equation} \label{Eq:7} 
\Delta B_{N} \sim C\mu _{0} m_{N} \rho _{N}^{1 / 2} /  d^{3/2},     
\end{equation} 
with $C$ a constant of order $\sim1/20$ obtained from integration of Eq. (\ref{Eq:6}) 
and $m_N $the nuclear magneton. For example, in the case of an organic system 
with proton density  $\rho_N\sim5\times10^{28} m^{-3}$ and assuming $d\sim200$ nm, 
we obtain $\Delta B_N\sim2.5$~nT, a value approaching the sensitivity limit of a 
room-temperature, diamond-based magnetometer~\cite{Taylor08,Maze08}. 

 We note that detection of the \textit{average} magnetization within the `active' 
volume --as opposed to magnetization \textit{fluctuations} --is conceivable if the 
contribution to the total field from spins outside this volume has been canceled.~\cite{Meriles05} 
In this case the nuclear field $\tilde{B}_{N} $ at the NV center site has the approximate 
value
\begin{equation} \label{Eq:8} 
\tilde{B}_{N} \sim D\mu _{0} m_{N} \rho _{N} P,       
\end{equation} 
independent of the sensor-surface distance. Here $P=m_NB_A/(2k_BT)$ is the nuclear 
Boltzmann polarization at temperature $T$ and $D$ is a constant of 
value $\sim1/6$. Comparing Eq. (\ref{Eq:7}) and (\ref{Eq:8}) we find the 
criterion for spin noise dominance, $d\le \left({k_{B} T / \left(m_{N} B_{A} 
\right)} \right)^{{2 / 
3} } \rho _{N}^{-{1 / 
3} } $. For example, if we take as a reference the case in which the protonated sample 
($\rho_N\sim5\times10^{28} m^{-3}$) has been polarized to the equivalent of a magnetic field $B_A$=10 
T at room temperature, we have $d\le $500 nm.

\section{Sensitivity limits} 
Having identified the source and magnitude of the field fluctuations at the sensor 
site, we now turn our attention to the general problem of using a quantum object---the 
NV center---to gather information on the fluctuating ensemble of sample spins. The 
average fluorescence in the presence of the nuclear field is calculated from. Combining 
Eqs. (\ref{Eq:2}) and (\ref{Eq:3}),
\begin{equation} \label{Eq:9} 
\left\langle Tr\left\{M\rho \right\}\right\rangle =\frac{1}{2} \alpha \left(1+\cos 
\theta \, \left\langle \cos \phi \right\rangle \right)+\frac{1}{2} \beta \left(1-
\cos \theta \, \left\langle \cos \phi \right\rangle \right),   
\end{equation} 
where brackets indicate expectation value and average over the different configurations 
of the nuclear system. In Eq. (\ref{Eq:9}) we assume that the nuclear magnetization 
is negligible and that $BN $(and therefore $\phi $) has a symmetric distribution 
(i.e., $\left\langle \phi ^{2k+1} \right\rangle =0$, $k$=1,2,3\dots ). By 
comparison with the case in which no nuclear field is present and in the limit $\left
\langle \phi ^{2} \right\rangle <1$, we define the \textit{signal} $S_A$ as
\begin{equation} \label{Eq:10} 
\begin{array}{l} {S_{A} \equiv \left\langle Tr\left\{M\rho \right\}\right\rangle 
_{\phi } -\left\langle Tr\left\{M\rho \right\}\right\rangle _{\phi =0} } \\ {\quad 
\, \, =\left(\alpha -\beta \right)\cos \theta \frac{\left(\left\langle \cos \phi 
\right\rangle -1\right)}{2} e^{-\left({\Delta t / T_{2e} } \right)^{\gamma } } \approx 
\frac{\left\langle \phi ^{2} \right\rangle }{4} \left(\beta -\alpha \right)\cos \theta 
\, e^{-\left({\Delta t / T_{2e} } \right)^{\gamma } } \, } \end{array}.  
\end{equation} 
In deriving Eq. (\ref{Eq:10}) we introduced the coherence decay of the sensor 
spin characterized by the relaxation time $T_{2e}$ and the exponent~\cite{Dutt07,Childress06} $\gamma\sim3$. 
Note that the presence of the nuclear field translates into a change of the NV center 
average fluorescence proportional to the nuclear-spin-induced phase variance. The 
`signal' amplitude also grows linearly with the difference between the average fluorescence 
in each of the two possible spin states and reaches a maximum value when the phase 
difference $\theta$ between the excitation and projection pulses is either zero 
or a multiple of $\pi$ (see Fig. 1). 

 In order to determine the limiting signal-to-noise ratio, we make use of the property $M^{k} 
=a^{k} {\left| 0 \right\rangle} {\left\langle 0 \right|} +b^{k} {\left| 1 \right
\rangle} {\left\langle 1 \right|} $ and that $\Delta a^{2} (\Delta b^{2} )=\alpha 
(\beta )$ for Poisson variables, to calculate the variance
\begin{equation} \label{Eq:11} 
\begin{array}{l} {\Delta M^{2} =\left\langle Tr\left\{M^{2} \rho \right\}\right\rangle 
-\left\langle Tr\left\{M\rho \right\}\right\rangle ^{2} =} \\ {\quad \; =\frac{1}{2} 
\left(\alpha +\beta \right)+\frac{1}{2} \left(\alpha -\beta \right)\cos \theta \, 
\left\langle \cos \phi \right\rangle e^{-\left(\Delta {t / T_{2e} } \right)^{\gamma } } +\frac{1}{4} 
\left(\alpha -\beta \right)^{2} \left(1-\cos ^{2} \theta \, \left\langle \cos \phi 
\right\rangle ^{2} e^{-\left(\Delta {t / T_{2e} } \right)^{\gamma } } \right)\; .} \end{array} 
\end{equation} 
The 
signal-to-noise ratio, $SNR$=$S_A/\Delta M$ is then
\begin{equation} \label{Eq:12} 
\begin{array}{l} 
SNR^{-2} =\frac{\Delta M^{2} }{S_{A}^{2} } = \\ 
=\frac{8e^{2\left(\Delta t/ T_{2e}  \right)^{\gamma } } }{\left\langle \phi ^{2} \right\rangle ^{2} } \left\{ \frac{\left(\alpha +\beta \right)+\left(\alpha -\beta \right)\cos\theta \, 
e^{-\left(\Delta t/T_{2e}  \right)^{\gamma } } }{\left(\alpha -\beta \right)^{2} \cos^{2} \theta } 
\right\}
+\frac{4e^{2\left(\Delta t/T_{2e} \right)^{\gamma } } }
{\left\langle \phi ^{2} \right\rangle ^{2} } 
\left\{\frac{1-\cos ^{2} \theta \left(1-\left\langle \phi ^{2} 
\right\rangle \right)e^{-\left(\Delta t/ T_{2e} \right)^{\gamma } } }{\cos ^{2} \theta } \right
\} 
\end{array} 
\end{equation} 
where, for simplicity, we have assumed $\left\langle \phi ^{2} \right\rangle <1$ and $\cos 
^{2} \theta \ne 0$. Note that in the limit $e^{-\left(\Delta {t / T_{2e} } \right)^{\gamma } 
} \sim 1$ the first (otherwise dominant) term can be cancelled if we choose $\theta=\pi$ and 
assume that ${\left| m_{s} =1 \right\rangle} $ is a `dark' state (i.e., $\beta =0$). 
The latter, however, is not always the case in practice because, as pointed above, 
we have $\alpha \cong 1.5\beta$ for direct NV spin detection. Therefore, we recast 
(\ref{Eq:12}) in the approximate form
\begin{equation} \label{Eq:13} 
SNR\approx 0.1\, \left\langle \phi ^{2} \right\rangle \sqrt{\alpha } \, e^{-\left(
\Delta {t / 
T_{2e} } \right)^{\gamma } } ,    
\end{equation} 
where we made use of the fact that in current experimental settings 
$\alpha \cong 1.5\times10^{-2} \ll1$.~\cite{Dutt07,Childress06} Hence, the optimal sensing time becomes 
a compromise between the increase in $SNR$ due to larger phase change 
$\sqrt{\left\langle \phi ^{2} \right\rangle }$ and the exponential decay due to 
decoherence. A similar sensitivity limit is obtained from the measurement of the 
signal fluctuation, as explained in Appendix A.

 Starting from (\ref{Eq:13}), we can obtain a numerical estimate of the total 
time $T$ necessary for $SNR=10$: At a distance $d\sim15$ nm 
from the surface, and for a densely protonated sample we use (\ref{Eq:7}) 
to find $\Delta B_N\sim110$ nT. For a sensing interval $\Delta t\sim40\mu$s $\ll T_{2e}\sim1$ 
ms we get $\left\langle \phi ^{2} \right\rangle \cong 0.6$, thus requiring $NA\sim2\times10^6$ 
repetitions and a total time $T_{p} \cong N_{A} \Delta t\cong 60\; s$ (note that 
in the present case $t_{prep} ,t_{read}\ll\Delta t$, see Fig. 1). This sensitivity 
limit could be improved enormously if single-shot read out was available. Some strategies 
toward single-shot readout have recently been proposed, such as better collection 
efficiency via coupling of NV center to a nano-photonic wave-guide~\cite{Babinec10} or readout enhanced by 
a nuclear spin memory.~\cite{Jiang09,Neumann10} In 
this last strategy, nearby nuclear spins (such as the nitrogen associated with the 
NV center or a $^{13}$C) are used to store the information regarding the state of the 
electronic NV spin, so that a given measurement can be repeated many times by mapping 
back the state of nuclear spin onto the electronic spin after each readout. With 
this technique, it is possible to further improve the $SNR$ although at the 
expense of a much longer readout time (approaching several milliseconds).

\section{Measurement of nuclear spin time correlations} 
In the previous section, we implicitly assume that the nuclear correlation time $T_{2n}$ 
is smaller than the single measurement time (in practice, of order $\sim\Delta t$) 
since successive measurements must be independent if they are to improve the $SNR$. 
However, the opposite regime $T_{2n}\gg\Delta t$ allows one to extract 
valuable spectroscopic information on the sample system. Intuitively, this is possible 
because, as nuclei evolve coherently from a random initial state, the correlation 
function---and thus the power spectrum---of sample spins can be determined from the 
statistics of successive, time-delayed measurements.~\cite{Davenport70} Consistent with the assumption 
that the nuclear system evolves unperturbed by the NV center and that it behaves 
as a classical magnetic field, we define the autocorrelation function 
\begin{equation} \label{Eq:14} 
K_{M} \left(\tau \right)=\left\langle Tr\left\{M\rho \left(\Delta t+\tau \right)\right\}\; Tr
\left\{M\rho \left(\Delta t\right)\right\} \right\rangle,    
\end{equation} 
with $\rho \left(t+\Delta t\right)$ denoting the density matrix that evolved under 
the action of the nuclear field between the times $t$ and $t+\Delta t$ (thus 
acquiring the phase $\phi(t+\Delta t)=\int _{t}^{t+\Delta t}\gamma _{e} B_{N} \left(t'
\right)dt' $). Note that since the phase acquisition takes a time $\Delta t$, me 
must restrict $\tau $ in Eq. (\ref{Eq:14}) and thereafter to $\tau \ge \Delta 
t$. Combining Eqs. (\ref{Eq:14}) and (\ref{Eq:2}) we find
\begin{equation} \label{Eq:15} 
K_{M} \left(\tau \right)=\frac{1}{4} \left(\alpha +\beta \right)^{2} +\frac{1}{4} 
\left(\alpha -\beta \right)^{2} \left(K_{c} \left(\tau \right)\cos ^{2} \theta +K_{s} 
\left(\tau \right)\sin ^{2} \theta +2\cos \theta \left\langle \cos \phi \right\rangle 
\right),
\end{equation} 
where $K_{c} \left(\tau \right)\equiv \left\langle \cos \phi \left(\tau +\Delta t
\right)\; \cos \phi \left(\Delta t\right)\right\rangle $ and $K_{s} \left(\tau \right)
\equiv \left\langle \sin \phi \left(\tau +\Delta t\right)\; \sin \phi \left(\Delta 
t\right)\right\rangle $. Using $\phi <1$, and choosing $\theta ={\left(2k+1\right)
\pi  / 
2} $, we recast Eq. (\ref{Eq:14}) in the simpler form
\begin{equation} \label{Eq:16} 
K_{M} \left(\tau \right)\cong \frac{1}{4} \left(\alpha +\beta \right)^{2} +\frac{1}{4} 
\left(\alpha -\beta \right)^{2} \left\langle \phi \left(\tau +\Delta t\right)\phi 
\left(\Delta t\right)\right\rangle . 
\end{equation} 
Eqs. (\ref{Eq:16}) and (\ref{Eq:10}) can be used to reconstruct the autocorrelation 
function $K_{\phi } \left(\tau \right)\equiv \left\langle \phi \left(\tau +\Delta 
t\right)\, \phi \left(\Delta t\right)\right\rangle $ and to determine the sample 
power spectral density of the phase $f$---here having the role of a stochastic 
variable describing a stationary random process---via the Wiener-Khintchine theorem~\cite{Davenport70} 
\begin{equation} 
\label{Eq:17} 
F_{\phi } \left(\nu \right)=\int _{-\infty }^{\infty }K_{\phi } \left(\tau \right)
\, e^{-i2\pi \nu \tau } d\tau  .     
\end{equation} 
Note that because of the finite phase acquisition time, Eq. (\ref{Eq:17}) is restricted to 
a bandwidth defined by the inverse of the separation between two successive measurements 
$\sim1/\Delta t$ (and has a central observation frequency determined by $n$/$\Delta t$, 
with $n$ representing the number of $\pi$-pulses within the contact time $\Delta t$).

\section{Imaging and spectroscopy of nuclear spins in biological systems}
 In this section we consider examples that illustrate some of the potential advantages---and 
limitations---of using the proposed technique to reconstruct an image or a local 
nuclear spin spectrum. In each of the simulations that follow we use a virtual `sample 
spin source' that we recreate in the most realistic way possible from results obtained 
with other techniques. For image reconstruction purposes we assume that the sensor---in 
the form of a cantilever-mounted scanning NV center---can be positioned relative 
to the sample surface with nanoscale precision. 

\begin{figure}[htb]
	\centering
		\includegraphics[scale=0.4]{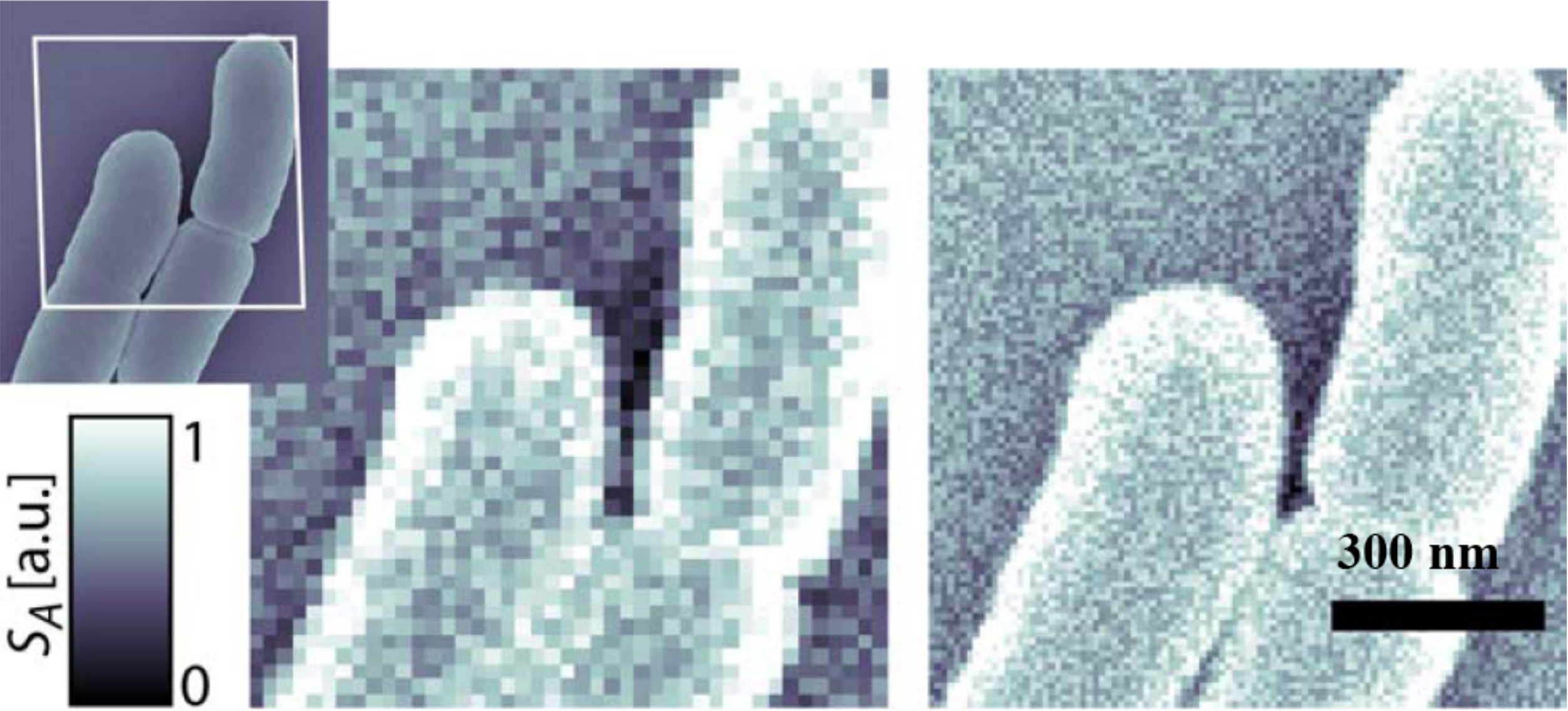}
	\caption{(Upper left insert) High-resolution SEM image of fixated \textit{E. 
Coli}. Brighter (darker) regions correlate with high (low) spin density. (Main images) 
`Raster scan' reconstructions of the corresponding 2D spin lattice.  The color code 
gauges the average NV center fluorescence as determined after $NA$ observations 
during which the spin alignment changes randomly (see Eq. (10)). The darker regions 
between and on the surface of the bacteria are artifacts resulting from artificial 
shadowing of the source SEM image. The tip-NV center distance and raster scan resolution 
is 30 nm (left) and 15 nm (right). The scale bar corresponds to 300 nm. Other parameters 
as listed in the text.}
	\label{fig:Fig2}
\end{figure}
We start by considering the SEM image of \textit{Escherichia coli} shown in the insert 
to Fig. 2. Specimens of this kind are usually fixated in a dry environment to preserve 
its morphology meaning that the color code in the image correlates with the spin 
density of protons. The two main images in Fig. 2 show the result of our simulations 
for which we considered the sample as a collection of classical, independent magnetic 
dipoles with a short correlation time (see below). A grey scale is used to indicate 
the average fluorescence of the NV center at each position ($S_A$ in Eq. \ref{Eq:10}). 
In our calculations, spins were distributed on a regular lattice with 1 nm separation 
and were given amplitude proportional to the local proton density (as implied by 
the SEM source image). The NV center distance to the sample surface was kept at $d=$15 
nm in one case (right) and 30 nm in the other (left). The evolution time $\Delta t$ was 
$40$ and $120\mu$s respectively and the number of measurements per image point was $N_A=4\times10^5$. 
The resulting time per pixel $T_p$ is 15 s (or 50 s) and the image time $T_i$ is 
estimated at 5 hs (or 4 hs) for a square of (500 nm)$^2$. We note that longer 
exposure times will be necessary if other, non-fundamental sources of noise are present; 
this scenario, however, is unlikely in an optimized confocal microscope where operation 
has been shown to be photon shot-noise limited.~\cite{Maze08}

One aspect of our example that deserves special consideration concerns the values 
assumed for the nuclear correlation and electron coherence times. First, we note 
that after fixation the system of Fig. 2 can be considered a solid with the result 
that the nuclear correlation time $T_N$ --assuming an external field $B_A$ stronger 
than the internuclear dipolar interaction-- is dictated by $T_{1N}$, the 
nuclear spin-lattice relaxation time. Because under realistic conditions $T_{1N}$ largely 
exceeds $\Delta t$, the time required for a raster scan of the sample grows to impractical 
values if the nuclear spin configuration at a given position must change randomly 
before the next measurement is carried out. Fortunately, there are ways to circumvent 
this problem, the simplest being to probe other points of the sample surface during 
the wait time. 

Attaining the longest coherence time in a NV center --exceeding 1 ms in isotopically 
depleted samples~\cite{Balasubramanian09}-- demands intercalating a $\pi$-pulse at the midpoint of the evolution 
time $\Delta t$.~\cite{Dutt07,Childress06} While, for simplicity, our calculations have obviated this 
need~\footnote{ Note that the inclusion of a $\pi$-pulse at the midpoint of the evolution 
interval $\Delta t$ can be easily accommodated by interpreting $T_{2e}$ as the homogeneous transverse-relaxation  time and by rewriting the accumulated phase in the form  
$\phi =\int _{0}^{\frac{\Delta t }2}\gamma_{e} B_{N}(t)dt -\int _{\frac{\Delta t}2}^{\Delta t}\gamma_{e} B_{N} (t)dt $ }, 
one immediate practical consequence is that a synchronous $\pi$-rotation must be 
applied on the sample spins if the net effect of the sample dipolar field on the 
sensor is to be preserved. In the presence of an auxiliary dc field $B_A\sim200$ 
gauss, the latter can be carried out via a resonant `radio-frequency' pulse (at $\sim1$ 
MHz).\footnote{With a duration of, for example, $\sim2\mu$s, the inversion pulse has 
no effect on the 13C ensemble surrounding the NV center thus preserving the `revivals' 
of the sensor spin echoes.\cite{Taylor08,Maze08} }

Although spin density mapping is the most basic form of imaging, it is ultimately 
the ability to introduce contrast between different soft tissues that separates MRI 
from other imaging technologies. Some of the contrast methodologies used in MRI find 
a natural extension in our detection strategy. For example, molecular diffusion away 
from the immediate vicinity of the sensor results in a shortening of the nuclear 
correlation time, which, with a suitable selection of $\Delta t$, can be exploited 
to make the time-averaged phase shift $\phi$ negligibly small.The latter 
is shown in Fig. 3 where we used an SEM image from the membrane of a red blood cell 
to encode the correlation time of spins on a uniformly dense lattice (i.e., spins 
in the void spaces of the SEM image were assigned a shorter nuclear correlation time). 
This example provides a rudimentary model for a `water-filled' membrane whose semi-rigid 
skeleton can be distinguished from the embedded fluid. 

\begin{figure}[htb]
	\centering
		\includegraphics[scale=0.4]{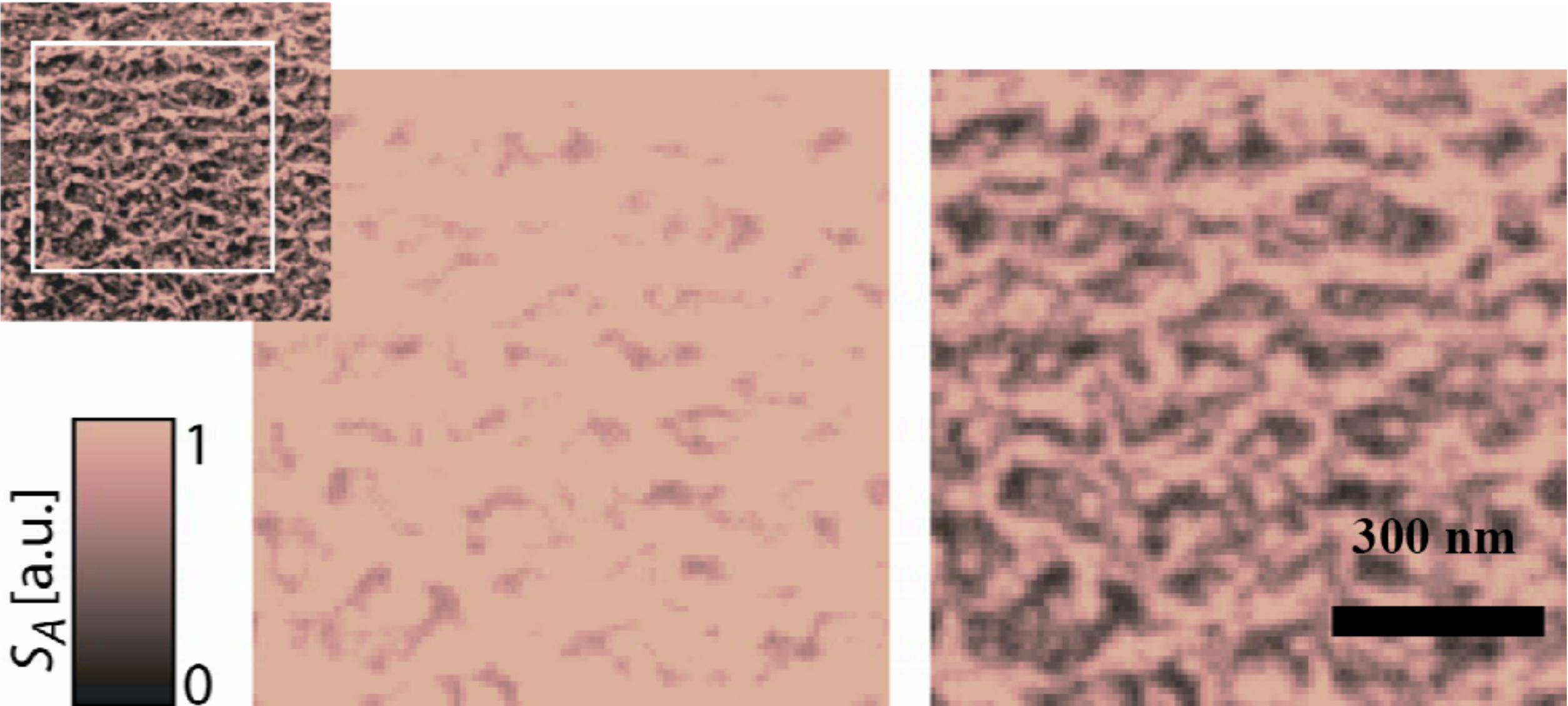}
	\caption{(Insert) SEM image of the membrane of a red blood cell. 
Void spaces become apparent only after dehydration and fixation. (Main images) Unlike 
Fig.~\ref{fig:Fig2}, the virtual 2D spin matrix is uniform (emulating the case 
of a `wet membrane'). This time the color scale of the source image was used to encode 
the local nuclear spin correlation time. In the example presented on the left, `mobile' 
regions (corresponding to dark regions in the source image) have a correlation time 
only 1.3 times shorter than the rest. The image on the right is based upon identical 
conditions except that the correlation time difference was three times greater. The 
scale bar corresponds to 300 nm.}
	\label{fig:Fig3}
\end{figure}

 In situations similar to that of Fig. 3, Eq. (\ref{Eq:17}) could be used, 
for example, to monitor diffusion processes. In this context, we note that one of 
the most important structural characteristics of the cell membrane is that it behaves 
like a two-dimensional liquid, i.e., its constituent molecules rapidly move about 
in the membrane plane. Therefore, one could imagine extensions of the basic pulse 
protocol to emulate their corresponding NMR counterparts (but with resolution on 
the tens of nanometers). In principle, a broad range of diffusion rates is within 
reach (because the probing time can be greatly enhanced if, after a given evolution 
period, the NV center coherence is stored in an adjacent $^{13}$C nucleus for future retrieval~\cite{Dutt07}). 
Studies of this kind may prove worthy, especially if we keep in mind that although 
the structure of plasma membranes is known to be inhomogeneous, the precise architecture 
of this important system still remains unclear.~\cite{Lillemeier06}

 In a different implementation where the auxiliary field $B_A$ points along an 
axis non-collinear with the crystal field one could rely on the above formalism to 
extract spectroscopic information from random nuclear spin coherences. An example 
is shown in Fig. 4 where we consider a set of (model) molecules with a 13 Hz heteronuclear 
(e.g., proton-phosphorous) $J-$coupling. In our simulation the auxiliary magnetic 
field $B_A$ is 20 gauss, the tip distance is 30 nm, and the system correlation 
time is 100 ms. Assuming the sensor at a fixed position in space, Fig. 4 shows the 
pulse sequence and resulting autocorrelation function and power spectral density. 
Implicit in this model is the idea that molecules tumble and move relative to each 
other so as to cancel inter- and intramolecular dipolar couplings without escaping 
the observation volume of the sensor during $\Delta t$. Our example mimics the conditions 
of `restricted diffusion' found, for example, within a cell membrane where molecules 
`hop' between adjacent, $\sim(100 nm)^3$ compartments on a time scale of several milliseconds.~\cite{Murase04}

\begin{figure}[htb]
	\centering
		\includegraphics[scale=0.38]{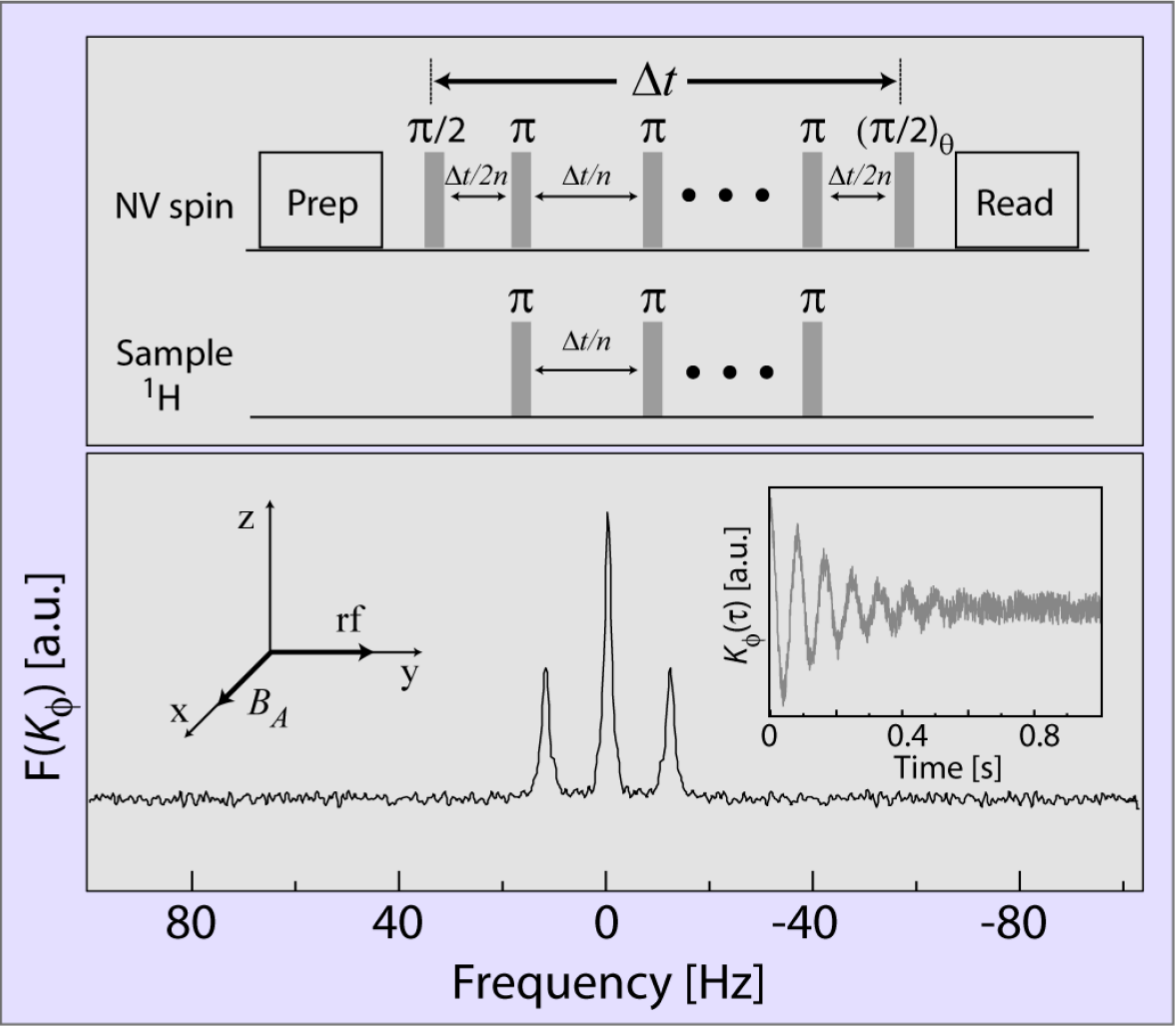}
	\caption{In this example the NV center repeatedly monitors a set 
of equivalent protons subject to a 6.5 Hz heteronuclear $J$-coupling with 
a second (invisible) spin-1/2 species. Depending on the alignment of the latter, 
protons precess with one of two possible frequencies. (Top) Schematics of the pulse 
sequence; $n$ denotes the number of p-pulses within the evolution interval 
$\Delta t$. (Bottom) Reconstructed correlation (insert) and corresponding spectral 
density. Note the factor 2 in the observed splitting (13 Hz), a direct consequence 
of having assumed $\left|\cos \theta \right|=1$ (`quadratic response'). In the simulation $n$=1, $d=$30 
nm and $\Delta t$=100 ms. The nuclear correlation time is 100 ms, and the number 
of single measurement pairs per point in the correlation curve is 4x105. The external 
magnetic field $B_A $is 5 mT and pulses acting on nuclear spins are assumed 
to be broadband so as to invert proton spins as well as the $J$-coupled species. 
Other conditions as listed in the text. }
	\label{fig:Fig4}
\end{figure}

  \section{Conclusion}

 While high-field MRI serves as a superb tool to probe the living world, achieving 
submicroscopic spatial resolution presently appears to be a goal exceedingly difficult. 
Indirect detection via NV centers in diamond provides an alternative platform that 
we examined by means of analytical and numerical calculations. We considered the 
particular case of a \textit{single} NV center interacting with a large number of 
nuclear spins, a condition that we described in semiclassical terms. When brought 
in close proximity to the sample surface, e.g. with the aid of a high-precision scanner, 
the NV center is selectively sensitive to field fluctuations induced by nuclear spins 
immediately adjacent to the sensor (even if the mean sample magnetization is negligible). 
The important practical consequence is that pre-polarization magnet, gradient coils, 
and fast-switching current amplifiers---today mandatory in a nuclear spin imaging 
experiment---are not requisites of this technology. 

 Our calculations show that simple Ramsey or spin-echo sequences are able to probe 
the nuclear spin system although the relative phase between pulses plays a crucial 
role. Under current experimental conditions, photon shot noise is the main source 
of error. We stress, however, that the sources of this limitation are not fundamental 
and that technical advances could lead to significant decrease in the imaging times. 

 When 
compared to other kinds of microscopies, several distinguishing features of NV center-based 
magnetometry emerge. For example, given the sharp dependence on the sensor-sample 
distance, detection is restricted to surface spins (with the result that careful 
sample preparation will be necessary when inner structures of a system are to be 
exposed). On the other hand, the same setup could be exploited to reconstruct 3D 
topographic maps that can then be used to enrich the information content of the images 
produced via the NV center fluorescence. Even if exposure times longer than those 
typical of other imaging schemes are necessary, diamond-based magnetometry has the 
potential to gauge changes in the dynamics and chemical composition of the sample, 
thus opening the door to various types of contrast. In particular, we have shown 
that, with an adequate protocol, one could probe molecular diffusion or reconstruct 
the low- or zero-field nuclear spin spectrum~\cite{McDermott02,Weitekamp83} with nanoscale spatial resolution. Finally, we note 
that most biochemical reactions are thermally-driven, stochastic processes that involve 
the crossing of a barrier or diffusion over some kind of potential energy surface. 
Therefore, the ability to conduct experiments in an open environment, at room temperatures 
can prove crucial to expose the dynamics of living systems in ways not possible with 
traditional magnetic resonance. For example, with spatial resolution of $\sim5$ nm---only 
slightly better than our target here---one could envision investigating the stepping 
of single molecular motors, a process that usually takes place in the tens of milliseconds 
range.

 \begin{acknowledgments}
We are thankful to Vik Bajaj for useful comments and to David Cowburn for 
their remarks on large portions of this manuscript. C.A.M. acknowledges support from 
the Research Corporation and from NSF. L.J. acknowledges support from Sherman Fairchild 
Fellowship. Work at Harvard is supported by the NSF, DARPA, and the Packard Foundation.\end{acknowledgments}

 \appendix

 \section{}

While 
through Eq. (\ref{Eq:10}) we monitor sample spins via changes in the \textit{average} number 
of photons emitted by the sensor, similar information can be obtained if we measure 
instead changes in the fluorescence \textit{variance. }This strategy closely mimics 
that already demonstrated in magnetic force microscopy7 and, within the framework 
presented above, appears as a `natural' alternate pathway. Starting from (\ref{Eq:11}) 
and neglecting for simplicity relaxation, a simple calculation shows that the signal $S_V$ in 
this case is given by
 \begin{equation}
  S_{V} \equiv \Delta M_{\phi }^{2} -\Delta M_{\phi =0}^{2} \cong \frac{1}{4} \cos 
\theta \, \left(\alpha -\beta \right)\left(\left(\alpha -\beta \right)\cos \theta 
-1\right)\left\langle \phi ^{2} \right\rangle. 
\label{eq:A1}
\end{equation}

To determine the limit uncertainty, we define the auxiliary operator $V
\equiv \left(M-\left\langle Tr\left\{M\rho \right\}\right\rangle \right)^{2} $ and 
calculate $\Delta V^{2} =\left\langle Tr\left\{\left(V-\left\langle Tr\left\{V\rho 
\right\}\right\rangle \right)^{2} \rho \right\}\right\rangle $. In the limit in which 
the shot noise is stronger than the spin noise, we find after a lengthy but straightforward 
calculation $\Delta V^{2} \cong \alpha $. Therefore, the signal to noise ratio is 
given in this case by 

 \begin{equation}
   SNR=\frac{S_{V} }{\Delta V} \cong 0.1\, \left\langle \phi ^{2} \right\rangle 
\sqrt{\alpha } , \end{equation} 
in agreement with (\ref{Eq:13}).

 We note that $S_A$ in Eq. (\ref{Eq:10}) --and thus $S_V$, Eq. 
(\ref{eq:A1})-- is insensitive to fluctuations of the nuclear field if the phase difference $\theta$ between 
the excitation and projection pulses is an odd multiple of $\pi/2$. In a way, 
this condition is counterintuitive because, in a sequence where the pulses are phase-shifted, 
the magnetometer responds linearly---not quadratically---to external fluctuations 
prompting the question as to whether higher sensitivity can be reached. 

Though in a different context, Wineland and collaborators have discussed similar 
problems extensively.~\cite{Itano93} Their 
work highlights the ambiguity that stems from the quantum character of the sensor 
via the concept of `quantum projection noise': When a \textit{single} two-level system 
probes a (non-fluctuating) magnetic field, maximum sensitivity comes at the price 
of complete uncertainty in the outcome of a measurement; reciprocally, when the measurement 
variance is zero, so is the sensitivity to external fields. Although in the present 
case the `signal' comes in the form of fluctuations of the magnetometer phase, this 
principle does play here an important (if more subtle) role. We can make it explicit 
by rewriting (\ref{Eq:11}) as
\begin{equation}\Delta M^{2} =\Delta M_{q}^{2} +\Delta M_{c}^{2} ,  \end{equation}
 where $\Delta M_{q}^{2} =\left\langle Tr\left\{M^{2} \rho \right\}-\left(Tr
\left\{M\rho \right\}\right)^{2} \right\rangle $ and $\Delta M_{c}^{2} =\left\langle 
\left(Tr\left\{M\rho \right\}\right)^{2} \right\rangle -\left\langle Tr\left\{M\rho 
\right\}\right\rangle ^{2} $. The first contribution measures the `quantum projection 
noise' or uncertainty in a population measurement of a single two level system; the 
second term corresponds to the nuclear-spin-noise-induced variance in a `classical', 
macroscopic-like sensor (where the average polarization can be determined from a 
single measurement). If, for simplicity, we consider in Eq. (\ref{Eq:2}) $a=1$ 
and $b=0$, we find after a simple calculation 
\begin{equation}
\Delta M_{q}^{2} =\frac{1}{4} \left(1-\left\langle \cos ^{2} \left(\phi +\theta 
\right)\right\rangle \right),
\end{equation}
 and
\begin{equation}
\Delta M_{c}^{2} =\frac{1}{4} \left(\left\langle \cos ^{2} \left(\phi +\theta 
\right)\right\rangle -\left\langle \cos \left(\phi +\theta \right)\right\rangle ^{2} 
\right). \end{equation}
 For the special case $\theta=(2k+1)\pi/2$ it follows 
 $\Delta M_{c}^{2} =\frac{1}{4} \left\langle \sin ^{2} \phi \right\rangle $ 
 and $\Delta M_{q}^{2} =\frac{1}{4} \left(1-\left\langle \sin ^{2} \phi \right\rangle \right)$ meaning that 
as we increase the amplitude of the external field fluctuations, the gain in the 
`classical' contribution to the variance is lost because of an equal but opposite 
change of the quantum projection noise. This is no longer the case when $\theta=k\pi$ thus 
leading to an observable change in the variance of the sensor fluorescence.

  \bibliographystyle{aipnum4-1}

\end{document}